\documentclass[useAMS,usenatbib]{mn2e}
\usepackage{bm, amsmath, amssymb, graphicx}
\usepackage[english]{babel}
\title[Kinematics of AM CVn]{Kinematics of the ultra-compact helium accretor AM Canum Venaticorum}
\author[G.\,H.\,A. Roelofs et al.]{G.\,H.\,A.~Roelofs,$^1$\thanks{E-mail: g.roelofs@astro.ru.nl} P.\,J.~Groot,$^1$ G.~Nelemans,$^1$ T.\,R.~Marsh$^2$ and D.~Steeghs$^3$\\
$^1$Department of Astrophysics, Radboud University, Toernooiveld 1, 6525 ED Nijmegen, The Netherlands\\
$^2$Department of Physics, University of Warwick, Coventry CV4 7AL, UK\\
$^3$Harvard-Smithsonian Center for Astrophysics, 60 Garden Street, Cambridge, MA 02318, USA}

\newcommand{\obj}{AM CVn}
\newcommand{\vel}{92$\pm$5 km/s}
\newcommand{\q}{$q=0.18\pm0.01$}
\newcommand{\e}{$e=0.04\pm0.01$}
\begin{document}
\maketitle

\begin{abstract}
We report on the results from a five-night campaign of high-speed spectroscopy of the 17-minute binary AM Canum Venaticorum, obtained with the 4.2-m William Herschel Telescope on La Palma.

We detect a kinematic feature that appears to be entirely analogous to the `central spike' known from the long-period, emission-line AM CVn stars GP Com, V396 Hya and SDSS J124058.03$-$015919.2, which has been attributed to the accreting white dwarf. Assuming that the feature indeed represents the projected velocity amplitude and phase of the accreting white dwarf we derive a mass ratio \q\ for AM CVn. This is significantly higher than the value found in previous, less direct measurements. We discuss the implications for AM CVn's evolutionary history and show that a helium star progenitor scenario is strongly favoured. We further discuss the implications for the interpretation of AM CVn's superhump behaviour, and for the detectability of its gravitational-wave signal with \emph{LISA}.

In addition we demonstrate a method for measuring the circularity or eccentricity of AM CVn's accretion disc, using stroboscopic Doppler tomography. We test the predictions of an eccentric, precessing disc that are based on AM CVn's observed superhump behaviour. We limit the effective eccentricity in the outermost part of the disc, where the resonances that drive the eccentricity are thought to occur, to \e, which is smaller than previous models indicated.

\end{abstract}


\section{Introduction}

AM Canum Venaticorum was found in a survey of faint, blue objects by \citet{hz}. It was observed to have peculiarly broad and shallow helium absorption lines, but no hydrogen \citep{greenstein}. The star was shown to be a possible ultra-compact white dwarf binary by \citet{smak67}, who discovered photometric variations on an 18-minute period; quickly thereafter, \citet{paczynski} noted that it would be a prime example of a binary whose evolution is expected to be governed by gravitational-wave radiation, and which could serve as an excellent test for the existence of gravitational radiation. The interpretation as an interacting binary analogous to the Cataclysmic Variables was first proposed by \citet{warner} upon their discovery of rapid flickering in the light-curve. The discussion as to the true orbital period of the system was finally put to rest by \citet{nsg} who discovered a kinematic `S-wave' feature in time-resolved spectra of AM CVn, thereby proving that the orbital period is 1028 seconds, while the main photometric signal at 1051 seconds is to be interpreted as a `superhump'.

Here we present phase-resolved spectra of AM CVn with significantly higher spectral resolution and signal-to-noise ratio than those previously used by \cite{nsg}. Moreover, our current data-set fully samples the proposed 13.37-h accretion disc precession period \citep{pattersonprec,skillman} for the first time, allowing for a characterisation of the spectroscopic appearance of the system as a function of orbital period, precession period, and superhump (beat) period.

\section{Observations and data reduction}

\begin{table}
\begin{center}
\begin{tabular}{l l r r}
\hline
Date		&UT		        &Exposures      &Typical\\
                &                       &(30--40\,s)      &seeing ('')\\
\hline
\hline
2005/04/20	&20:52--05:06	        &523            &0.8\\
2005/04/21	&22:46--05:15	        &114            &0.8\\
2005/04/22	&20:41--05:04	        &585            &1.1\\
2005/04/23	&21:03--04:57	        &370            &1.4\\
2005/04/24	&20:35--05:01	        &567            &0.8\\
\hline
\end{tabular}
\caption{Summary of our observations.}
\label{observations}
\end{center}
\end{table}

We obtained phase-resolved spectroscopy of \obj\ on 20, 21, 22, 23 and 24 March 2005 with the William Herschel Telescope (WHT) and the ISIS spectrograph. The observations consist of 2159 spectra taken with the 1200B grating, covering $\sim$4300--5100\AA; the exposure time was 30--40 seconds depending on sky transparency and airmass. The sky varied from clear to opaque due to high clouds, but the median seeing was quite good, about $0.8''$, giving an effective resolution of about 0.7\,\AA\ or 45 km/s.

The chosen slit width was dependent on the seeing and transparency, and varied from $1.0''-1.4''$. The detector was the standard EEV12 chip in the spectrograph's blue arm, windowed and binned by a factor of two in the spatial direction to increase the read-out speed. The read-out speed was kept `low' to minimize the read-out noise. Each night, an average bias frame was created from 20 individual bias exposures, and a normalised flatfield frame was constructed from 20 incandescent lamp flatfields.

All spectra were extracted using the IRAF implementation of optimal (variance-weighted) extraction. The read-out noise and photon gain, necessary for the extraction, were calculated from the bias and flatfield frames, respectively. He\-Ne\-Ar arc exposures were taken every hour during the night to correct for instrumental flexure; all dispersion solutions were interpolated between the two arc exposures nearest in time. Arcs were taken before and after every rotator adjustment needed to keep the slit at near-parallactic angle. In each arc exposure a total of about 40 arc lines could be fitted well with a Legendre polynomial of order 4 and 0.04\,\AA\ root-mean-square residuals. The root-mean-square residuals due to flexure are estimated to be smaller than this; the total arc drift was measured to be 0.5 \AA\ peak-to-peak over an entire night. All spectra were transformed to the heliocentric rest-frame prior to analysis.

The average spectrum was corrected for instrumental response using spectrophotometric standard star Feige 34.

A summary of all observations is given in table \ref{observations}.

\section{Results}

\subsection{Average and phase-binned spectra}
\label{averagespectrum}

\begin{figure}
\centering
\includegraphics[angle=270,width=84mm]{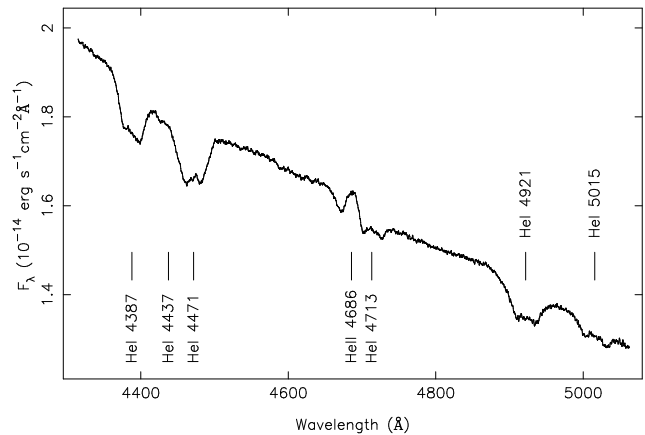}
\caption{Average spectrum of AM CVn. The wavelength-dependent instrumental response has been corrected for, but the overall received flux is strongly affected by clouds; no attempt has been made to correct for this.}
\label{average}
\end{figure}

The grand-average spectrum of AM CVn is shown in figure \ref{average}. It shows several broad neutral helium absorption lines seen before in AM CVn's spectrum, and clear emission of \mbox{He\,{\sc ii}} 4686. The absorption wings extend to about 1300 km/s (half width at half minimum). See table \ref{eqw} for a list of equivalent widths for the strongest absorption lines.

\begin{table}
\begin{center}
\begin{tabular}{l r}
\hline
Line		                                        &Equivalent width (\AA)\\
\hline
\hline
\mbox{He\,{\sc i}} 4387                                 &$1.58\pm 0.02$\\
\mbox{He\,{\sc i}} 4471 (+\mbox{He\,{\sc i}} 4437)      &$2.84\pm 0.02$\\
\mbox{He\,{\sc ii}} 4686 (+\mbox{He\,{\sc i}} 4713)    &$0.76\pm 0.04$\\
\mbox{He\,{\sc i}} 4921                                 &$2.60\pm 0.05$\\
\mbox{He\,{\sc i}} 5015                                 &$0.73\pm 0.02$\\
\hline
\end{tabular}
\caption{Equivalent widths of the detected accretion disc absorption lines, including estimated errors, in the average spectrum of AM CVn (figure \ref{average}).}
\label{eqw}
\end{center}
\end{table}

\begin{figure}
\centering
\includegraphics[angle=270,width=84mm]{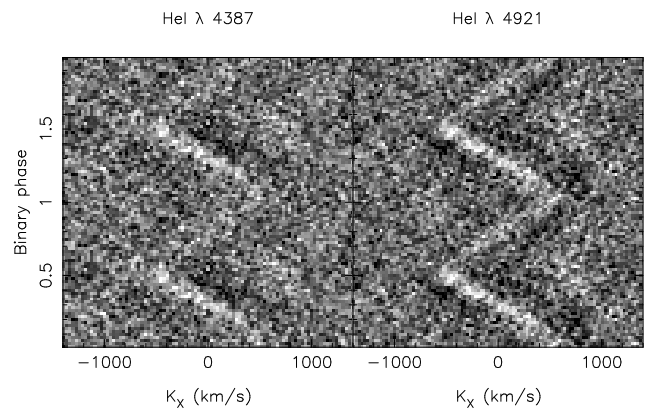}
\caption{Phase-binned, average-subtracted spectrum of AM CVn around the \mbox{He\,{\sc i}} 4387 and \mbox{He\,{\sc i}} 4921 lines.}
\label{othertrails}
\end{figure}

The phase-binned (trailed) spectrum reveals several additional, weaker features that are too smeared out in the average spectrum to be detected. See figures \ref{othertrails} and \ref{spiketrail}. Among these are the `S-wave' emission features in several lines, already found by \citet{nsg}. It includes the metal line \mbox{Mg\,{\sc ii}} 4481; already present but not mentioned in \citet{nsg}. Two very weak S-wave features flank the (also very weak) \mbox{He\,{\sc i}} 5048 line, which we identify with \mbox{Si\,{\sc ii}} 5041 \& 5056.

The most remarkable is a weak and narrow emission feature slightly offset to the red of \mbox{He\,{\sc i}} 4471. It is clearly offset in phase from the presumed bright spot feature, and has a much lower amplitude (see figure \ref{spiketrail}). As such it has everything in common with the `central spike' feature that was first detected in GP Com \citep{smak75} and has since become a common (though poorly understood) feature in long-period, emission-line AM CVn stars \citep{mtr, roelofs}. As in the aforementioned systems, the phase of the low-velocity component (relative to the bright spot) is consistent with the expected phase of the accreting white dwarf, and its relatively low velocity amplitude is difficult to reconcile with any emission site in the accretion disc. It is also intrinsically redshifted through some as yet unknown mechanism (cf.\ \citealt{trm99,lmr}). Based on these three properties, we interpret the feature as the perfect analogue of the central spike in GP Com, V396 Hya, and SDSS J124058.03$-$015919.2.

We see no evidence for a central spike in any of the other helium lines. However, from \citet{lmr} we see that the central spike is strongest in the \mbox{He\,{\sc i}} 4471 line in GP Com, and it is clear that we are barely detecting this line's spike in AM CVn. Our non-detection of a spike in the other helium lines is thus plausibly a matter of signal-to-noise.

\subsection{Doppler tomography}
\label{tomography}

\subsubsection{The central spike}

\begin{figure}
\centering
\includegraphics[angle=270,width=84mm]{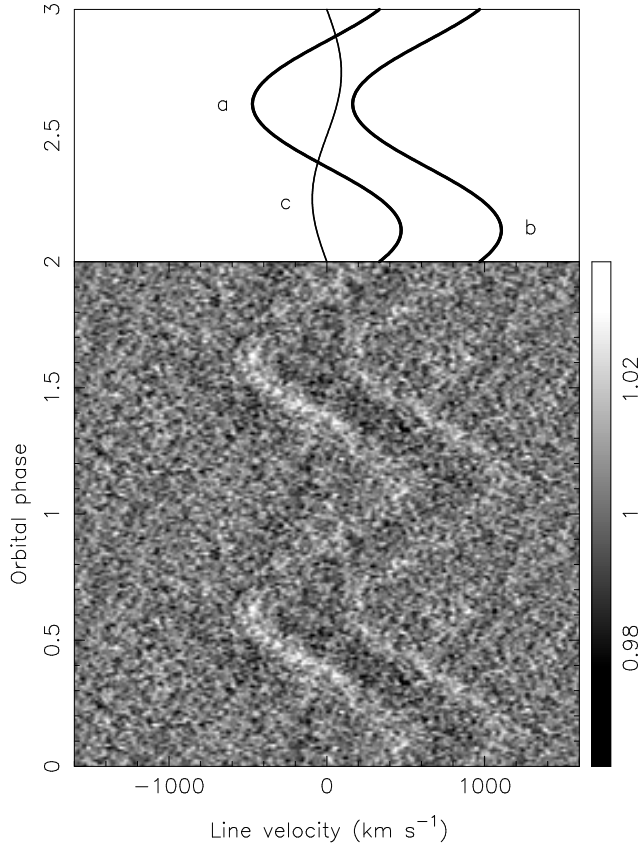}
\caption{Phase-binned, average-subtracted spectrum of AM CVn around the \mbox{He\,{\sc i}} 4471 line. The bright spot trails around zero velocity (component \textsf{a}), in phase with the bright spot from the nearby \mbox{Mg\,{\sc ii}} 4481 line (\textsf{b}). In addition there is a weak component moving around 40 km/s with a semi-amplitude of \vel, and offset in phase from the bright spot (\textsf{c}). We propose that this is the `central spike' of the \mbox{He\,{\sc i}} 4471 line.}
\label{spiketrail}
\end{figure}

The weak, sinusoidal emission feature we interpret as the central spike in the \mbox{He\,{\sc i}} 4471 line can be made more tangible in a linear back-projection Doppler tomogram \citep{dopplermapping}. We begin by phase-binning our 2159 spectra in $\sim$200 bins using the known orbital period $P_\mathrm{orb}=1028.7322$\,s \citep{harvey,skillman}. The two bright spots close to the spike -- from the \mbox{He\,{\sc i}} 4471 line itself and from the nearby \mbox{Mg\,{\sc ii}} 4481 line, see figure \ref{spiketrail} -- are masked out to prevent them from dominating the (back-projected) spectra. Each phase-binned spectrum is then normalised by fitting a polynomial and dividing the spectrum by it, and finally all phase bins are divided by the column-averaged spectrum so that we are left with the components of the spectrum that vary on the orbital period, apart from the bright spots which were masked out earlier.

\begin{figure}
\centering
\includegraphics[angle=270,width=84mm]{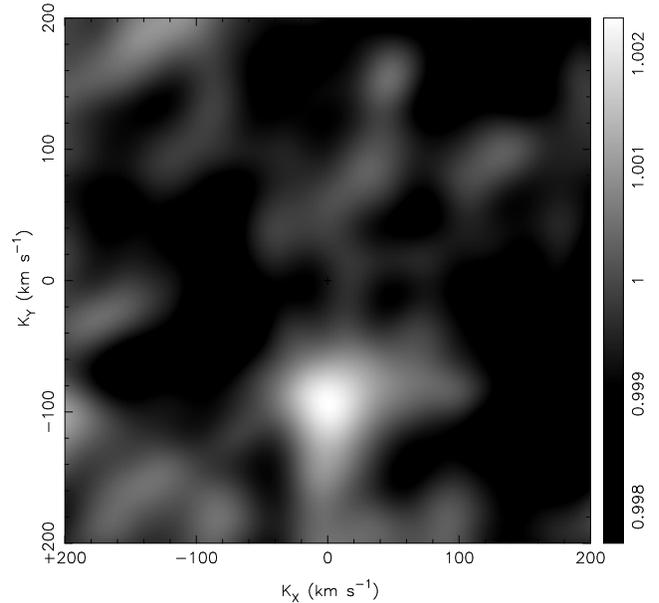}
\caption{Linear back-projection Doppler tomogram of the redshifted, low-velocity central spike. Its velocity amplitude comes out at \vel. Its phase has been aligned with the negative $K_Y$ axis, the conventional zero-phase of the accreting star.}
\label{spikeback}
\end{figure}

The next step is determining the central spike's rest wavelength or, equivalently, its intrinsic (rest) velocity $\gamma$ relative to the \mbox{He\,{\sc i}} 4471 rest wavelength. We proceed in the same manner as in \cite{roelofs}, that is, we make Doppler tomograms for a range of trial wavelengths and determine the wavelength at which the spectra auto-correlate best. As in \cite{roelofs}, the height of the presumed central spike nicely peaks at a certain wavelength, which we take as the rest wavelength of the central spike. See table \ref{dopplerfeatures} and the corresponding Doppler tomogram in figure \ref{spikeback}.

The final step is determining the central spike's velocity amplitude and phase, and their associated errors. We employ the bootstrap method used in \cite{roelofsaw}. In this Monte Carlo process we make a large number ($10^3$) of Doppler tomograms with the recipe described above, using a random selection of 2159 spectra taken from our data set, and allowing for replacement. We determine the emission centroid of the central spike in each tomogram, make a 2-D histogram of all the $K_X$ and $K_Y$ values obtained, and fit a 2-D Gaussian to this distribution. The width of this Gaussian is taken as a measure for the error on the central spike's coordinates in $K_X$, $K_Y$ space. See table \ref{dopplerfeatures} for the results.

\subsubsection{The bright spot}

We follow the same procedure as for the central spike, but now masking out only the bright spot of the \mbox{Mg\,{\sc ii}} 4481 line. We again construct an ensemble of $\sim$$10^3$ Doppler tomograms via the bootstrap method in order to be able to estimate the errors on the resulting data points. Table \ref{dopplerfeatures} lists the rest wavelengths, velocity amplitudes and phases as well as their errors, of both the bright spot and the central spike of the \mbox{He\,{\sc i}} 4471 line. The central spike has been aligned with the negative $K_Y$ axis, which coincides with the conventional zero-phase of the accreting white dwarf.

\begin{table}
\begin{center}
\begin{tabular}{l r r r}
\hline
Feature		&$\gamma$       &$K_Y$          &$K_X$\\
                &(km/s)         &(km/s)         &(km/s)\\
\hline
\hline
Bright spot	&$0\pm5$        &$330\pm8$      &$-336\pm8$\\
Central spike	&$40\pm4$       &$-92\pm4$       &$0\pm4$\\
\hline
\end{tabular}
\caption{Velocities of the bright spot and central spike components of the \mbox{He\,{\sc i}} 4471 line, after aligning the central spike with the negative $K_Y$ axis.}
\label{dopplerfeatures}
\end{center}
\end{table}

\begin{figure}
\centering
\includegraphics[angle=270,width=84mm]{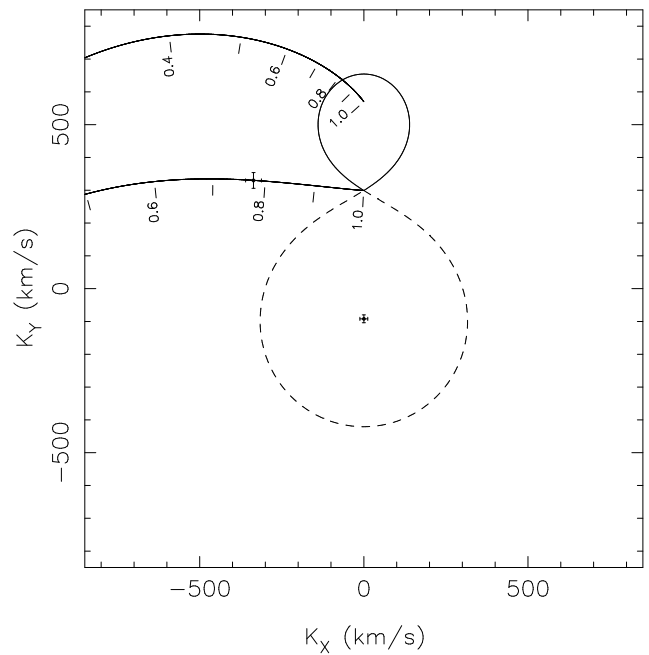}
\caption{The measured central spike and bright spot velocities, with the associated 3-$\sigma$ error bars. Overplotted is the best-fitting binary of mass ratio $q=0.18$, with the ballistic accretion stream starting at the inner Lagrange point and passing through the bright spot. The upper solid line starting within the secondary star represents the Keplerian accretion disc velocities along the ballistic stream trajectory. The labels represent the accretion disc radius as a fraction of $R_{L1}$.}
\label{doppdata}
\end{figure}

Figure \ref{doppdata} shows the velocities of the central spike and the bright spot, with errors, to show their relative phases and amplitudes, after aligning the central spike with the negative $K_Y$ axis. Looking ahead at the next section, we also plot the velocities of a free-falling accretion stream, as well as the Keplerian velocities along the trajectory of the ballistic accretion stream, for a binary of mass ratio $q=M_2/M_1=0.18$.

\subsection{The mass ratio of AM CVn}
\label{massratiosection}

\begin{figure*}
\centering
\includegraphics[angle=270,width=\textwidth]{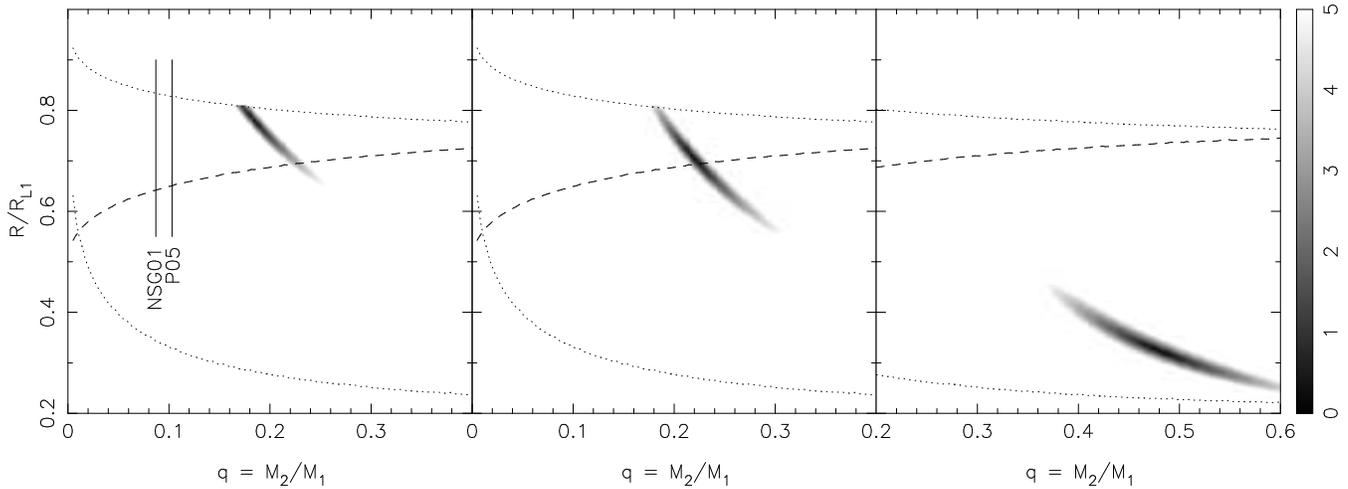}
\caption{Mass ratio $q$ and effective accretion disc radius $R$ of AM CVn based on the observed velocities and phases of the central spike and the bright spot. Left: purely ballistic stream velocities; right: purely Keplerian disc velocities; centre: mixing of 80\% stream and 20\% disc velocities in the bright spot. The grey-scale shows the exclusion confidence levels, in standard deviations. The upper and lower dotted lines indicate the edge of the primary Roche lobe and the circularisation radius, respectively, while the dashed line shows the $3:1$ resonance radius. The labels NSG01 and P05 indicate previous mass ratio estimates based on AM CVn's superhump behaviour \citep{nsg,patterson}.}
\label{massratio}
\end{figure*}

Using the velocity amplitudes and phases of the central spike and the bright spot, we can constrain the mass ratio and the effective accretion disc radius in AM CVn. To this end we solve the equation of motion for a free-falling stream of matter through the inner Lagrangian point, based on the results of \cite{lubowshu}, and we see whether the resulting accretion stream and/or accretion disc velocities and phases at the stream--disc impact point match with the measured values. The results are displayed in figure \ref{massratio}.

Keplerian velocities in the bright spot imply very small effective accretion disc radii ($R\sim 0.3R_{L1}$, with $R_{L1}$ the distance from the centre of the accretor to the inner Lagrange point), and very large mass ratios ($q\sim0.5$). In particular, the disc would not come anywhere near the $3:1$ resonance radius. This resonance is commonly thought to drive the superhump behaviour that is observed in outbursting dwarf novae as well as nova-like systems such as AM CVn \citep{whitehurst91,hirose}.

For purely ballistic stream velocities in the bright spot, the best-fitting accretion disc radius is $5-10\%$ smaller than the maximum accretion disc radius that can be contained within the primary star's Roche lobe. In this case, the disc would clearly extend past the $3:1$ resonance radius. If we assume purely ballistic stream velocities, we obtain a mass ratio \q. This constitutes a lower limit.

If we allow for mixing between the ballistic stream and the Keplerian disc velocities, the mass ratio could be slightly larger than this. See figure \ref{massratio}. The maximum mass ratio for which the bright spot would remain outside the $3:1$ resonance radius is $q\leq 0.22\pm0.01$, which corresponds to a (best-fitting) mix of about 80\% stream and 20\% disc velocities. A mass ratio of $q\lesssim 0.25$ is commonly quoted as the requirement for a binary to be able to excite its $3:1$ resonance and develop the corresponding superhumps \citep{whitehurst88,whitehurst91,hirose}.

Two values for the mass ratio $q$ from the literature are overplotted in figure \ref{massratio}. Both these values are derived from the fractional superhump period excess $\epsilon\equiv\left(P_\mathrm{sh}-P_\mathrm{orb}\right)/P_\mathrm{orb} = 0.0216$ observed in AM CVn (e.g.\ \citealt{skillman}), combined with theoretical-numerical or empirical relations between $\epsilon$ and $q$. The mass ratio derived from AM CVn's kinematics is significantly larger than these predictions. The present data indicate a lower limit of \q, compared with $q=0.10$ as obtained from the latest empirical $\epsilon(q)$ relation for hydrogen-rich Cataclysmic Variables by \citet{patterson} (P05). The value $q=0.087$ found by \citet{nsg} (NSG01), from fits to numerical accretion disc simulations as taken from \citet{warnerbook}, deviates even more. In the remainder of this paper, we shall conservatively use \q, which corresponds to pure stream velocities in the bright spot.

\subsection{Stroboscopic Doppler tomography}

The permanent superhump phenomenon, exhibited by AM CVn, is usually explained in terms of a precession of a tidally deformed (i.e., non-circular) accretion disc. After exactly one orbital cycle, the orientation of the two stars with respect to the observer is restored, but during the orbit the eccentric disc has advanced slightly so that it takes slightly more than one orbital period to restore the orientation of the two stars with respect to the accretion disc. This beat period is then assumed to be responsible for the main photometric signal, which is usually observed to be a few percent longer than the orbital period.

The model requires that there be a precessing accretion disc with sustained eccentricity. Furthermore, \citet{skillman}, among others, find the superhump cycle to be extremely stable over many thousands of orbital periods, with the superhump period always within 0.1 s of 1051.2 s and the superhump waveform remaining very stable throughout the years\footnote{Note that the superhump waveform is double-humped; most of the power is actually in the first harmonic at 525.6 seconds.}. This implies that the shape of the disc is (semi-)stationary rather than chaotic. It further implies that the superhump phase will drift by at most 0.04 of a cycle over our $\sim$100-hour observing baseline, if we adopt their average superhump period of 1051.2 seconds. These two results open up the possibility of combining the large number of superhump cycles in our data-set to reconstruct the appearance of the system as a function of superhump phase. 

The application of Doppler tomography as an instrument for testing the eccentric disc model is as follows. As the secondary star goes around the slowly precessing and non-circular disc, the effective radius at which the accretion stream impacts the disc will vary. Since the bright spot is caused by the free-falling stream of matter from the inner Lagrange point crashing into the edge of the accretion disc, a variation in effective accretion disc radius will lead to differences in the gas velocities of both the stream and the disc matter at the impact point. Observationally, we then expect the bright spot to show excursions in velocity space, as a function of the relative orientation of the eccentric disc and the secondary star, i.e., as a function of superhump phase.

In order to measure these excursions we must divide our spectra in a number of bins corresponding to different superhump phases, and subsequently phase-fold the spectra contained in each bin on the orbital period. The method bears resemblance to the `stroboscopic Doppler tomography' method employed by \citet{stroboscopy} to disentangle the orbital and white dwarf spin periods in the magnetic accretor FO Aqr, except that we now have the orbital period of the binary and the precession period of the accretion disc.

We argue that this is in fact quite a sensitive method for measuring variations in the effective radius of the accretion disc. If we consider the simplest case of an eccentric, elliptical accretion disc, a simple estimate for the ratio of the projected apastron and periastron velocity amplitudes $v_\mathrm{a,p}$ as a function of the eccentricity $e$ is
\begin{equation}
\frac{v_\mathrm{p}}{v_\mathrm{a}} \approx \frac{1+e}{1-e} \approx 1 + 2e, \quad e\ll 1.
\label{ratio}
\end{equation}
A bright spot of average velocity amplitude $\overline{v}\simeq 450$ km/s, combined with a reasonable spectral resolution and signal-to-noise ratio that allow the bright spot to be measured to 10 km/s accuracy, gives an eccentricity sensitivity of $e\sim 0.01$. In practice, the excursion of the bright spot in a Doppler tomogram will be larger and the eccentricity sensitivity will correspondingly be better than this.

\begin{figure}
\centering
\includegraphics[angle=270,width=84mm]{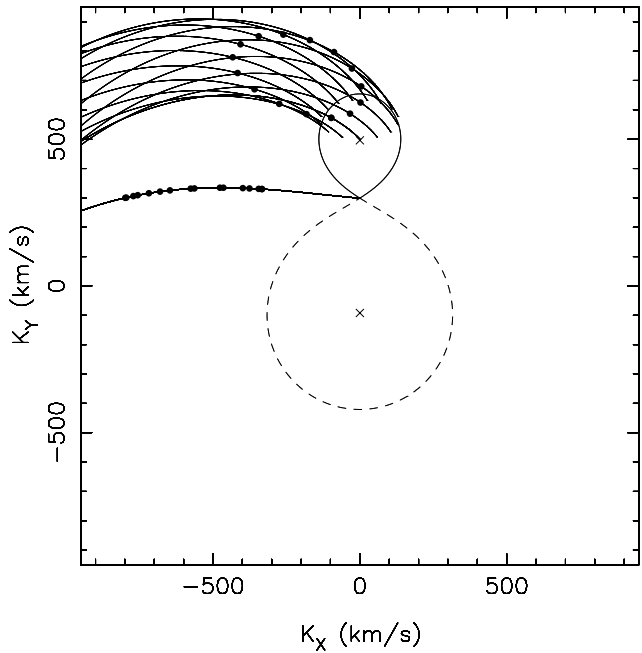}
\caption{Simulated stroboscopic Doppler tomogram for a binary of mass ratio $q=0.18$, accretion disc semi-major axis $R=0.65R_{L1}$, and accretion disc eccentricity $e=0.2$. The elliptical group of spots represent Keplerian accretion disc velocities in the bright spot, the group sticking out to the left represent ballistic accretion stream velocities, as a function of superhump phase (one full cycle shown).}
\label{doppsim}
\end{figure}

As an example we show in figure \ref{doppsim}, by solving numerically the equation of motion for a ballistic test mass in the binary potential, how we expect the bright spot to vary as a function of superhump phase, in a binary of mass ratio $q=0.18$, with an elliptical accretion disc of semi-major axis $R=0.65R_{L1}$ and eccentricity $e=0.2$. This eccentricity value is chosen based on the simple observational model of \cite{pattersonprec} and the numerical models of \cite{simpsonwood}. The bright spot is expected to show excursions of several hundreds of km/s, indicating that eccentricities of this magnitude should be easily observable with our spectral resolution.

In our simple model we have assumed a somewhat sharp edge to the accretion disc as in e.g.\ \citet{armitagelivio}, such that the gradual changes in pressure along the ballistic stream do not significantly alter the depth of the stream's impact into the disc. In order to explore the possible effects of such variations, consider the pressure of the accretion stream $p_\mathrm{s}$,
\begin{equation}
p_\mathrm{s} = p_\mathrm{static} + p_\mathrm{dynamic} \approx p_\mathrm{dynamic} = \frac{1}{2}\rho v^2
\end{equation}
where $\rho$ and $v$ are the density and velocity of a fluid element along the stream, respectively, and $p_\mathrm{static}$ can be neglected since $k_BT\ll m_\mathrm{He}v^2$ in the free-fall regime (by definition, one could say), where $v$ is of the order $10^2$\,km/s and the temperature $T$ of the stream, set by the surface temperature of the donor star, is of the order $10^4$\,K. If we assume a steady ballistic flow of gas, where the scale height of matter in the flow remains fairly constant in the region just outside the disc \citep{lubowshu2} since the matter does not have time to maintain hydrostatic equilibrium perpendicular to the flow, then conservation of mass gives
\begin{equation}
\rho \propto v^{-1} \qquad \mathrm{(along\ streamline)}
\end{equation}
so that
\begin{equation}
p_\mathrm{s} \propto v \qquad \mathrm{(along\ streamline)}
\end{equation}

The pressure along the accretion stream thus increases linearly with velocity. In case of a `soft' edge to the accretion disc (that is, a slowly increasing disc pressure $p_\mathrm{d}$ from the disc edge inwards) this could have the effect of \emph{increasing} the ratio of observed periastron and apastron bright spot velocities for given $e$, as the stream impacts slightly deeper into the disc rim at periastron due to its higher pressure than it does at apastron. Thus if one employs the radius at which the bright spot occurs as the effective \emph{in situ} accretion disc radius, and one allows the disc edge to be soft, an eccentricity measurement based on the bright spot excursion amplitude could lead to an overestimate of $e$, and thus to an \emph{upper limit} for $e$.

\subsection{Eccentricity of the accretion disc of AM CVn: the bright spot}

\begin{figure}
\centering
\includegraphics[angle=270,width=84mm]{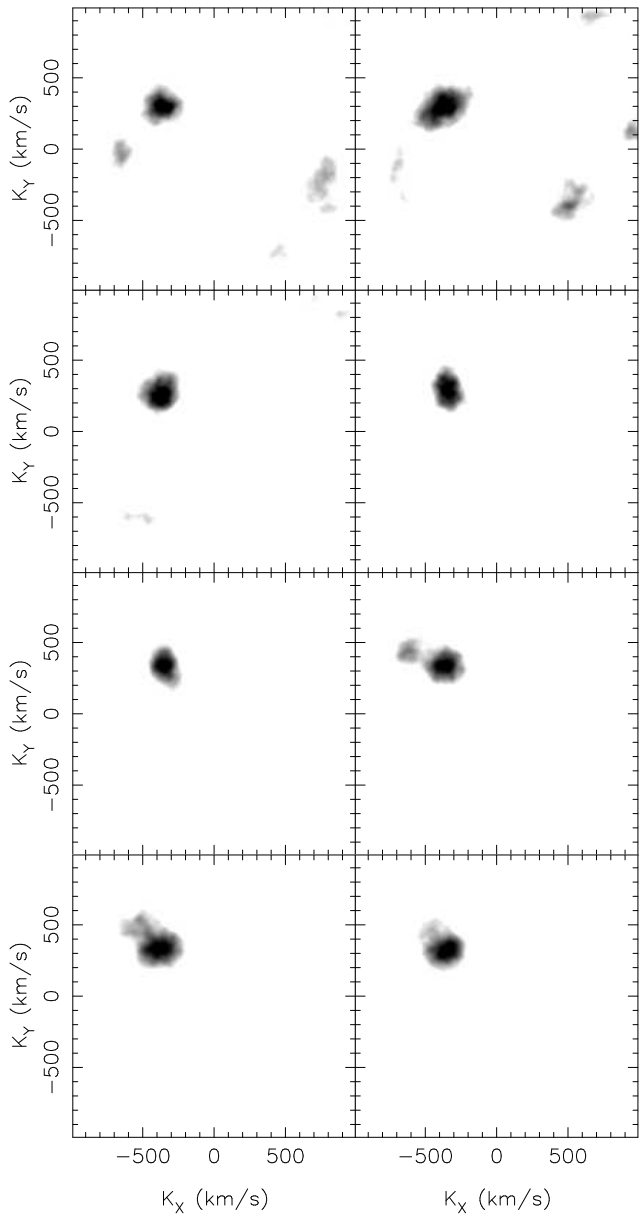}
\caption{Doppler tomograms showing the bright spot as a function of superhump phase. One full cycle is shown from left to right, top to bottom. The bright spot is visible at all superhump phases, and shows no large excursions in velocity space.}
\label{precspotmaps}
\end{figure}

\begin{figure*}
\centering
\includegraphics[angle=270,width=\textwidth]{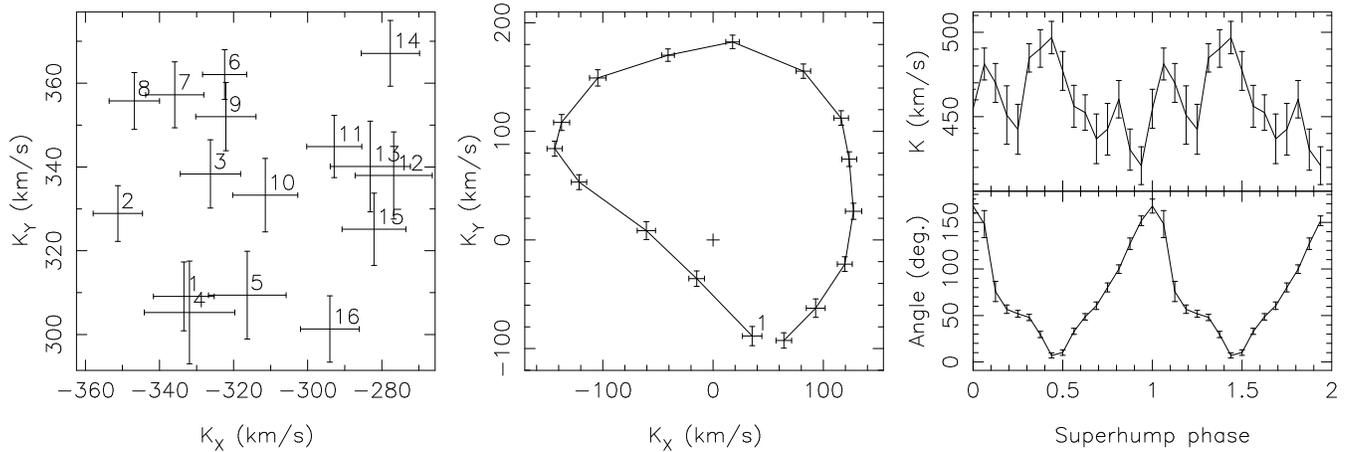}
\caption{The accretion disc as a function of superhump phase. Left: the measured bright spot velocities in 16 superhump phase bins. Middle: the measured \mbox{He\,{\sc ii}} 4686 emission centroids, again in 16 phase bins. Right: the bright spot velocity amplitude, together with the angle (in spatial coordinates) between the bright spot and the \mbox{He\,{\sc ii}} 4686 emission centroid, as a function of superhump phase. In this last panel we assume approximately stream velocities in the bright spot and Keplerian velocities for the \mbox{He\,{\sc ii}} 4686 emission.}
\label{precession}
\end{figure*}

Figures \ref{precspotmaps} and \ref{precession} show the results from our stroboscopic Doppler tomograms. In fig.\ \ref{precspotmaps} we plot eight Doppler tomograms that together cover one full superhump period. In order to improve on the signal-to-noise ratio of the individual maps we combined the data (in velocity space) of four lines that show a clear bright spot in the trailed spectra, namely \mbox{He\,{\sc i}} 4387, \mbox{He\,{\sc i}} 4471, \mbox{Mg\,{\sc ii}} 4481 and \mbox{He\,{\sc i}} 4921 (figures \ref{othertrails} \& \ref{spiketrail}). Figure \ref{precession} shows the measured excursions of the bright spot as a function of superhump phase, for 16 superhump phase bins, connected by a line, where the bright spot centroids have been determined in the same way as in section \ref{tomography}. Here, the 16 phase bins oversample the superhump cycle by a factor of two, so that each bin contains 1/8th of the data as in figure \ref{precspotmaps}.

Based on the observed bright spot excursion amplitude of less than 100 km/s (see figure \ref{precession}), we constrain the effective eccentricity of the outermost part of the accretion disc to \e, where we define the `effective' eccentricity as that which would cause a simple elliptical disc to give rise to the same bright spot excursion amplitude in velocity space. The error on this value can be rather small because the method of measurement is insensitive to the accretion disc's semi-major axis (or average radius), and to whether the bright spot represents stream or disc velocities (see also figure~\ref{doppsim}). In case the bright spot velocity is a \emph{variable} mix of the disc and stream velocities, the derived eccentricity can be an overestimate.

Despite the rather large error bars in the left panel of figure \ref{precession}, at least relative to the small movement of the bright spot, the disc appears to be more complex than the simple ellipse which it is often pictured to be in explaining the superhump phenomenon. Detailed numerical simulations, e.g.\ by \cite{simpsonwood}, show that the accretion disc in a permanent superhump system like AM CVn is in fact most likely of irregular shape. Not only that, the disc is also predicted to change shape during a superhump period, only to return to its original shape after one full superhump cycle -- therefore, the disc could \emph{in principle} be exhibiting strong changes in radius at the other side of the accreting star, where the mass stream cannot be used to probe its effective radius. The changes in effective radius of the accretion disc \emph{might} thus be larger than is revealed by the excursions of the bright spot.

\subsection{Eccentricity of the accretion disc of AM CVn: integrated disc profile}

\begin{figure*}
\centering
\includegraphics[angle=270,width=\textwidth]{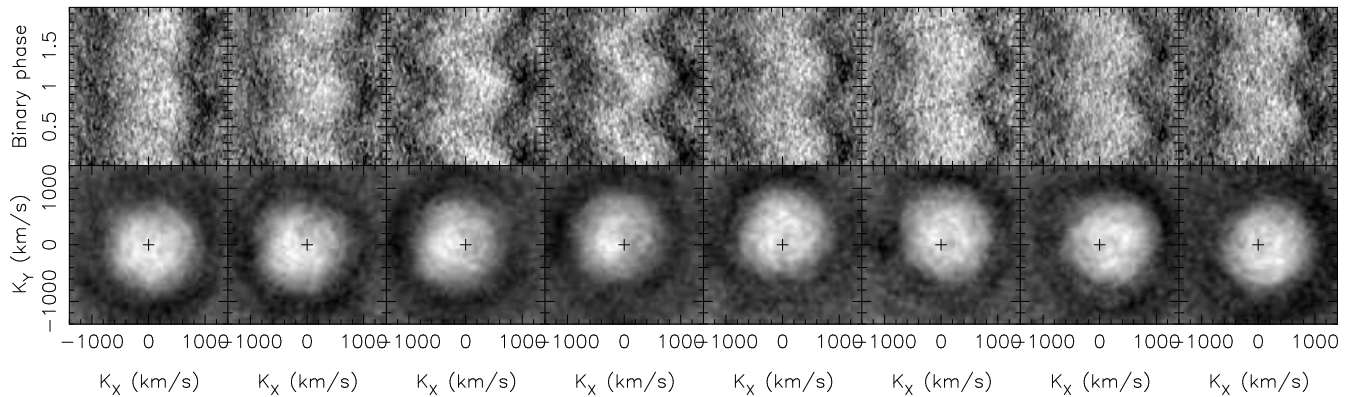}
\caption{Trailed spectra and linear back-projection Doppler tomograms of the \mbox{He\,{\sc ii}} 4686 line as a function of superhump phase. One full superhump cycle is shown from left to right.}
\label{heliumemission}
\end{figure*}

Figure \ref{heliumemission} shows trailed spectra and Doppler tomograms of \mbox{He\,{\sc ii}} 4686 as a function of superhump phase. The \mbox{He\,{\sc ii}} 4686 line was chosen because it clearly shows emission in the average spectrum (fig.\ \ref{average}), which is commonly used as a tracer of locally enhanced dissipation in accretion discs, for instance in searches for spiral shocks in outbursting dwarf novae. A sinusoid with the phase and amplitude of the central spike -- i.e.\ the presumed motion of the accretor -- has been shifted out of the spectra to leave the net kinematic signal of the matter in the disc orbiting the accretor. The ionised helium emission profile is clearly variable on the superhump period: the emission is asymmetric and the degree of asymmetry is changing (see also figure \ref{precession}). These general features agree with the predictions of \cite{simpsonwood}.

It can furthermore be deduced that the \mbox{He\,{\sc ii}} 4686 emission is strongest where the effective radius of the disc is smallest.
The right panel in figure \ref{precession} shows both the bright spot velocity amplitude and the angle between the bright spot and the centroid of \mbox{He\,{\sc ii}} 4686 emission, as a function of superhump phase. This angle is given in spatial coordinates -- for the transformation of velocity into spatial coordinates we assume roughly ballistic stream velocities for the bright spot based on the results of section \ref{massratiosection}, and roughly Keplerian disc velocities for the \mbox{He\,{\sc ii}} 4686 emission. Despite the rather large error bars, there appears to be a trend whereby the maximum bright spot velocity amplitude (which should correspond to the smallest effective disc radius) corresponds to alignment of the bright spot with the \mbox{He\,{\sc ii}} emission centroid, while the minimum bright spot velocity (i.e.\ largest effective disc radius) corresponds to anti-alignment.

These findings again agree with the numerical simulations of \citet{simpsonwood}. Their analysis of the changes in internal energy in the disc as a function of superhump cycle, too, showed that energy production was highest where the disc radius was smallest. This may not be that surprising since streamlines of matter in the disc will be squeezed more closely together towards a region of smaller disc radius, causing extra viscous dissipation. We may tentatively conclude that cooling proceeds on a timescale less than the orbital period of the matter in the disc, which is a few minutes.

\section{Discussion}

\subsection{A central spike in AM CVn?}
The mysterious central spike has so far been observed in long-period, low mass transfer rate AM CVn stars, where the accreting white dwarf dominates the optical flux. It is thought to originate on (or very close to) the surface of the accreting white dwarf, since the spike perfectly tracks the expected movement of the accreting white dwarf relative to the bright spot in the accretion disc. In addition, there is no accretion disc component that is expected to move at such low velocities, and the phase of the spike relative to the bright spot does not match the phase of the secondary. See \cite{lmr} for a detailed analysis of the central spike in GP Com.

The kinematic feature observed in the \mbox{He\,{\sc i}} 4471 line in \obj, as presented here, has all the characteristics for being the perfect analogy to the central spike in GP Com \citep{lmr}, V396 Hya (Steeghs et al.\ in preparation) and SDSS J124058.03$-$015919.2 \citep{roelofs}: (1) it is intrinsically redshifted; (2) it agrees in amplitude with the expected velocity amplitude of the accreting white dwarf (and not with any other binary component's velocity amplitude); and (3) it agrees in phase with the expected phase of the accreting white dwarf, relative to the bright spot. We therefore conclude that it is \emph{very} likely that it is the central spike.

The mass ratio for AM CVn implied by the central spike is significantly higher than previously thought, although there has been a study by \citet{pearson} suggesting, in fact, a mass ratio $q\sim0.19-0.25$ based on a model in which the secondary star is moderately magnetic. The motivation for that study was to try to reconcile the mass ratio with the $K$-velocity of the \mbox{He\,{\sc ii}} 4686 emission line measured (but discarded) by \citet{nsg}, which suggested a mass ratio in this regime. It is clear from this work however, in particular figure \ref{heliumemission}, that the measured $K$-velocities of the \mbox{He\,{\sc ii}} 4686 emission indeed \emph{cannot} be used reliably as an indicator for the motion of the accretor, since the apparent kinematic signal could easily be dominated by the eccentricity of the disc rather than the actual motion of the primary.

\subsection{Implications for AM CVn's formation channel}

One of the long-standing questions regarding the AM CVn stars is how they are formed. Theoretically, they can be formed in both double-degenerate and single-degenerate configurations: Roche-lobe overflow from a white dwarf onto another (more massive) white dwarf (e.g.\ \citealt{nelemans}), or from a helium-burning star onto a white dwarf \citep{ibentutukov}. A third suggested formation channel is Roche-lobe overflow from an evolved main-sequence star onto a white dwarf \citep{podsi}, where mass transfer starts by the time hydrogen core burning ends. This latter scenario, however, has the problem of leaving significant amounts of hydrogen, while our high-S/N average and phase-binned spectra show exclusively helium and metals. For the other two formation channels, the main discriminator is the mass of the secondary star: in the helium-star channel, one expects a relatively hot and massive, only partially degenerate secondary, while in the double-degenerate channel, one expects a `cold' degenerate donor star.

The significantly higher mass ratio found for AM CVn in this paper has important implications for the nature of the donor star, and hence for AM CVn's formation channel, as it suggests a relatively massive donor. If we combine the orbital period with the mass--radius relation for a relatively `cold' (core temperature $T_c \leq 10^6$\,K), degenerate helium white dwarf as used in \citet{nelemans} and refined recently in \citet{deloye}, we get a secondary mass $M_2\approx 0.035M_{\sun}$ and hence, through our minimum mass ratio of \q, a maximum primary mass of only $M_1=0.19M_{\sun}$. This is \emph{very} low; the maximum combined mass of $M_1+M_2=0.22M_{\sun}$ would, for instance, be significantly lower than the average value $M_1+M_2=0.8M_{\sun}$ found in detached WD--WD binaries (e.g.\ \citealt{nelemansspy}). This casts doubt on the possible white dwarf (WD) nature of the donor star, and thereby on the double-degenerate formation channel for AM CVn, unless the white dwarf donor was still very hot upon Roche-lobe overflow -- that is, unless mass transfer started shortly after the second common-envelope phase.

Concurrent with the measurement of AM CVn's mass ratio presented here, it was discovered that AM CVn's distance is larger than expected, through an \emph{HST/FGS} parallax measurement ($\pi=1.65\pm0.30$ mas, Roelofs et al.\ in preparation). This means a smaller absolute magnitude $M_V$ and thus, since we expect the accretion disc to dominate the optical flux, a higher accretion luminosity.

In order to link the observed $M_V$ to the mass of the secondary $M_2$, we proceed in the same way as in \cite{deloye}. We assume conservative mass transfer that is driven entirely by loss of angular momentum due to the emission of gravitational waves and, since accretion proceeds via an accretion disc, we assume that all angular momentum carried by the transferred matter is fed back to the orbit. We then have
\begin{equation}
\frac{\dot{M_2}}{M_2} = \frac{\dot J}{J} \frac{2}{\zeta_2+5/3-2q}
\label{mdotmeq}
\end{equation}
where $J$ is the orbital angular momentum, $\dot J$ is the orbital angular momentum loss rate
\begin{equation}
\frac{\dot J}{J} = -\frac{32}{5}\frac{G^3}{c^5} \frac{M_1 M_2 (M_1+M_2)}{a^4}
\end{equation}
due to the emission of gravitational waves \citep{landau}, and
\begin{equation}
\zeta_2 \equiv \frac{\mathrm{d} \log R_2}{\mathrm{d} \log M_2}.
\end{equation}
We take $\zeta_2=-0.06$ for $M_2 < 0.2M_\odot$ as in \citet{nelemans}, based on evolutionary calculations for a mass-losing helium-burning star by \citet{tf}. Clearly $\zeta_2 = 0$ around the point where helium burning stops, while it may be somewhat closer to $\zeta_2=-0.19$ along the semi-degenerate part of the track for a secondary that is more helium depleted, as found by \citet{savonije}. This represents a $\lesssim$10\% uncertainty in eq.\ (\ref{mdotmeq}).

The last step is to calculate the bolometric luminosity $L$, for which we use
\begin{equation}
L = \frac{1}{2} \dot{M_2} \big( \Phi(L_1) - \Phi(R_1) \big)
\label{Lacc}
\end{equation}
where $\Phi$ is the common Roche potential, at the inner Lagrange point $L_1$ and the surface of the accretor $R_1$. This states that the matter in the disc stays virialised during the accretion process, which should hold for approximately Keplerian particle orbits.

\begin{figure}
\centering
\includegraphics[angle=270,width=84mm]{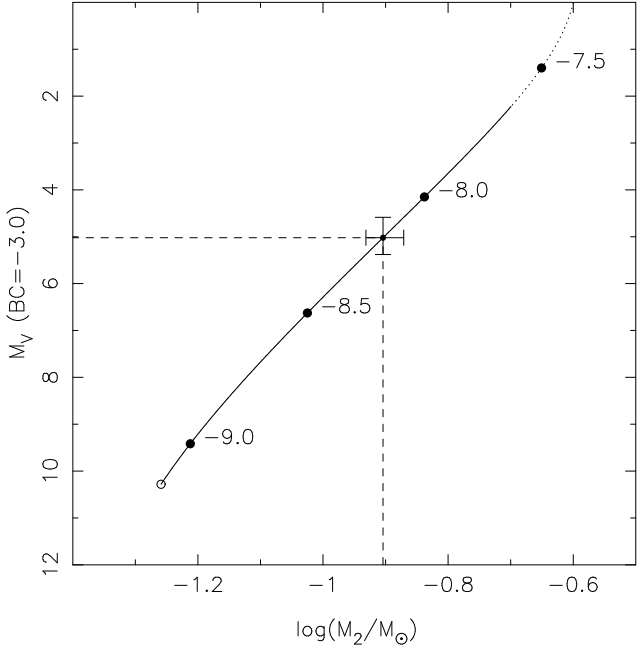}
\caption{Relation between $M_2$ and $M_V$ for AM CVn. Our measurement of the absolute magnitude $M_V$ and its implications for the secondary mass $M_2$ are indicated by the dashed line. The open circle represents the minimum secondary mass for which the system does not eclipse (assuming a disc that extends to the $3:1$ resonance radius); the dotted line, towards larger $M_2$ and smaller $M_V$, shows the helium-burning regime where we expect our assumptions regarding the mass-radius relation to break down ($>$10\% error). The labelled dots indicate $\log\dot{M_2}$ in $M_\odot$/yr.}
\label{mdotmplot}
\end{figure}

Figure \ref{mdotmplot} shows $M_V$ as a function of $M_2$ for $q=0.18$, $\zeta_2=-0.06$ and a bolometric correction $\mathrm{BC}=-3.0$ (see below). The lower limit to $M_2$ is given by the requirement that the system does not eclipse. The data point $M_V = 5.0\pm0.4$ as determined from our aforementioned \emph{HST} parallax is overplotted, which yields a secondary mass $M_2=0.125\pm0.012\,M_\odot$ and, via \q, a primary mass $M_1 = 0.68\pm0.06\,M_\odot$. The corresponding mass transfer rate is $\dot{M_2} = 6.7^{+1.9}_{-1.3}\cdot 10^{-9}M_\odot$/yr, and the inclination of the binary comes out at $i=43\pm2^\circ$. Table \ref{parameters} lists the system parameters for AM CVn in convenient tabular form.

\begin{table}
\begin{center}
\begin{tabular}{l l}
\hline
Parameter               &Value\\
\hline
\hline
$P_\mathrm{orb}$ (s)    &$1028.7322\pm0.0003$ \citep{skillman}\\
$q$                     &$0.18\pm0.01$\\
$M_1 (M_\odot)$         &$0.68\pm0.06$\\
$M_2 (M_\odot)$         &$0.125\pm0.012$\\[.5ex]
$\dot{M_2} (M_\odot/\mathrm{yr})$         &$6.7^{+1.9}_{-1.3}\cdot10^{-9}$\\[.5ex]
$i$ (degrees)           &$43\pm2$\\[.5ex]
$d$ (pc)                &$606^{+135}_{-93}$ (Roelofs et al.\ in prep.)\\
\hline
\end{tabular}
\caption{The collection of system parameters for AM CVn.}
\label{parameters}
\end{center}
\end{table}

Our results have been corrected for the geometric effect that the `apparent absolute magnitude' \citep{warnerbook} of an accretion disc depends on the inclination. A factor
\begin{equation}
\Delta M_V(i) = -2.5 \log \left(\frac{\cos i}{1/2}\right)
\end{equation}
has been taken into account as the correction from apparent absolute magnitude to absolute magnitude, where the factor $1/2$ represents the direction-averaged fraction of the disc area seen by the observer, $\langle\cos i\rangle$. We neglect the (unknown) effect of limb darkening.

In converting the bolometric luminosity $M_\mathrm{bol}$ obtained from equation (\ref{Lacc}) to an absolute visual magnitude $M_V$, the bolometric correction is of some importance. AM CVn's spectral energy distribution from the far-UV to the optical (Roelofs et al.\ in preparation; see also \citealt{nasser}) indicates that it is dominated by a blackbody of $\simeq$$30,000$\,K. This corresponds to a BC of $-3.0$, with an estimated error of $0.3$. If we calculate the minimum temperature the accretion disc must have in order to be able to radiate away the required luminosity (for any given $M_{1,2}$ we know the maximum radiating surface of the disc), we find $T_\mathrm{disc}\gtrsim 30,000$\,K for accretion rates $\dot{M_2} \ge 4\cdot 10^{-9} M_\odot$/yr. The BC we use is thus self-consistent with the accretion rate we derive; in particular, it seems unlikely that we are overestimating the mass transfer rate due to an overestimate of the BC.

\begin{figure}
\centering
\includegraphics[angle=270,width=84mm]{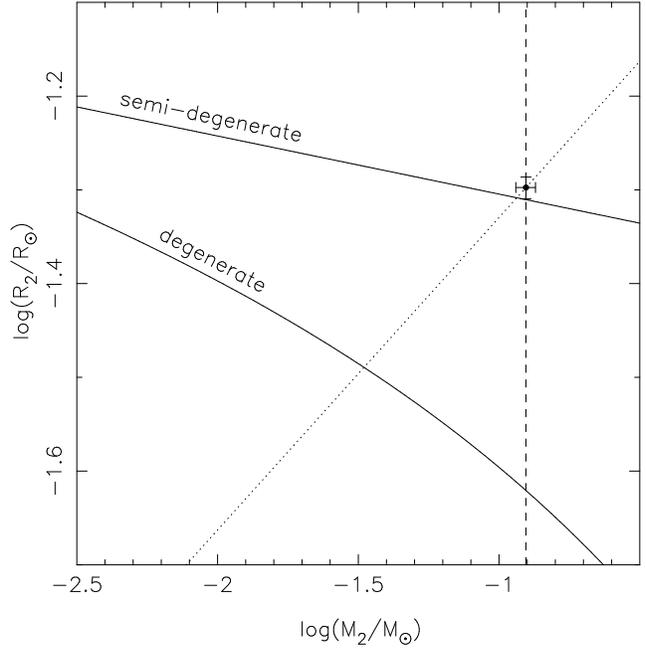}
\caption{Relation between $M_2$ and $R_2$ for AM CVn. The upper and lower solid line represent the evolutionary tracks for semi-degenerate helium star and for cold white dwarf secondaries, respectively \citep{nelemans,tf}. The dotted line is the requirement that the secondary star fills its Roche lobe. The dashed line represents our measurement of $M_2$. The data favour a semi-degenerate helium star secondary, rather than a white dwarf.}
\label{mrplot}
\end{figure}

Summarising, we have two independent pieces of evidence suggesting a relatively massive donor star -- the large mass ratio and the large luminosity. If we compare our results to evolutionary models for helium star and white dwarf secondaries as given by \citet{nelemans}, we see that our derived secondary mass is compatible with the secondary being the semi-degenerate core of a formerly helium burning star, while a `cold' degenerate donor star is ruled out. See figure \ref{mrplot}. The secondary mass $M_2=0.125M_\odot$ is close to the turn-off mass for helium burning, which lies at $M_2\sim 0.16 M_\odot$ \citep{savonije,tf}. We conclude that it is very likely that AM CVn formed through an evolutionary channel in which the donor star was, upon the start of Roche-lobe overflow, still burning helium. These observations thus provide the first observational evidence that the helium star evolutionary channel contributes to the AM CVn population.

\subsection{Implications for AM CVn's gravitational wave signal}

We expect the orbit of AM CVn to be circular, so that the gravitational wave polarisation amplitudes $A_+$ and $A_\times$ at twice the orbital frequency, $f$, become (e.g.\ \citealt{timpano})
\begin{align}
A_+ &= 2\frac{(G\mathcal{M})^{5/3}}{c^4d} \left(\pi f\right)^{2/3} \left(1+\cos^2i\right)\\
A_\times &= -4\frac{(G\mathcal{M})^{5/3}}{c^4d} \left(\pi f\right)^{2/3} \cos i
\end{align}
where $\mathcal{M}=(M_1M_2)^{3/5}/(M_1+M_2)^{1/5}$ is the so-called chirp mass,
$i$ is the inclination of the binary, and $d$ its distance. Defining the strain amplitude
\begin{equation}
h=\left(\frac{1}{2}\left(A^2_{+} + A^2_\times\right)\right)^{1/2}
\end{equation}
and converting to more convenient solar units, one finds
\begin{align}
h &= 2.84 \cdot 10^{-22}\sqrt{\cos^4i+ 6\cos^2i + 1}\nonumber\\
&\phantom{=}\times\left(\frac{\mathcal{M}}{M_\odot}\right)^{5/3} \left(\frac{P_\mathrm{orb}}{\mathrm{1\,hr}}\right)^{-2/3} \left(\frac{d}{\mathrm{1\,kpc}}\right)^{-1}
\end{align}

\begin{figure}
\centering
\includegraphics[angle=270,width=84mm]{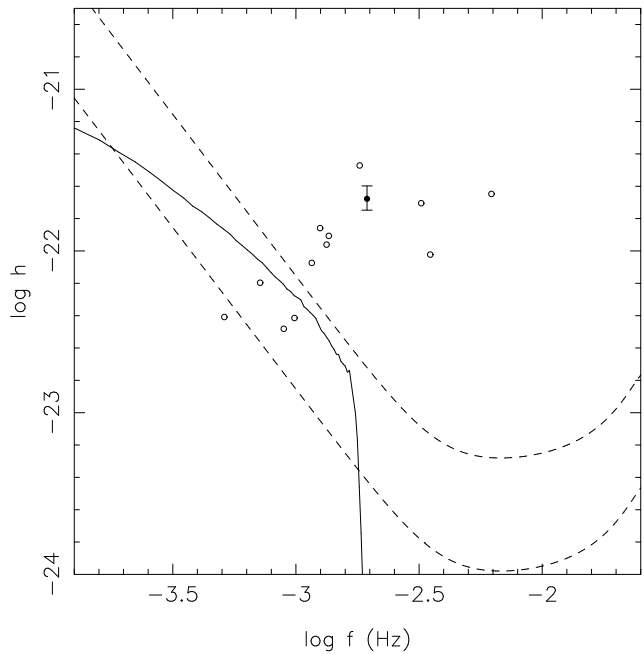}
\caption{Gravitational-wave strain amplitude of AM CVn (solid circle). The upper and lower dashed lines show the design sensitivities of \emph{LISA} for a S/N of 5 and 1, respectively, in one year of data-collecting \citep{larson}. The solid line is a population synthesis prediction for the confusion-limited Galactic background \citep{npy}. The open circles represent `known' AM CVn stars that are as yet not well-constrained in terms of either distance or component masses, or both.}
\label{gwr}
\end{figure}

Filling in the numbers derived above gives $h=2.1^{+0.4}_{-0.3}\cdot 10^{-22}$. Figure \ref{gwr} shows the estimated strain amplitudes and frequencies of the `known' AM CVn stars. AM CVn itself is the first potential \emph{LISA} source which is sufficiently well-constrained that we can plot error bars with some confidence, and we see that it stands out significantly above both the instrumental design sensitivity and the estimated confusion-limited Galactic background due to, mainly, detached WD--WD binaries.

\subsection{The superhump excess--mass ratio relation}

It is understood from numerical simulations, and it is observed in outbursting dwarf novae and nova-likes, that there exists a relation between the superhump period excess $\epsilon$ and the mass ratio $q$ of the binary. See \citet{patterson}, and references therein, for an overview including the very latest empirical work on the $\epsilon(q)$ relation. In short, they conclude that for hydrogen-rich systems,
\begin{equation}
\epsilon(q) = 0.18 q + 0.29 q^2
\label{epsilon}
\end{equation}
For AM CVn, this results in $q=0.10$ based on the photometry of \citet{skillman}. Although it is unknown what the intrinsic spread in $q(\epsilon)$ is, it appears from figure 9 in \citet{patterson} to be quite low, no more than about $0.01$. This means that the $q(\epsilon)$ for AM CVn deviates significantly from our minimum kinematic mass ratio \q.

Of course, the superhump excess $\epsilon$ may depend on more parameters than just the mass ratio $q$; it is quite conceivable that the helium nature of AM CVn may be of influence. A recent study by \citet{goodchild} indicates, for instance, quite a significant dependency of $\epsilon$ on the thickness of the accretion disc, where thicker discs lead to smaller superhump period excesses for a given mass ratio. If this effect should be at play, our results would suggest a physically thicker disc in AM CVn compared to discs in hydrogen-rich systems of the same mass ratio.

Although AM CVn represents only a single data point, it seems advisable to apply the empirical superhump excess--mass ratio relation with caution for such ultra-compact binaries. An obvious test would be to obtain the mass ratio in HP Librae, the other known helium nova-like, and check whether it, too, shows a discrepancy with relation (\ref{epsilon}).

\subsection{Size and eccentricity of the accretion disc}

\begin{figure}
\centering
\includegraphics[angle=270,width=84mm]{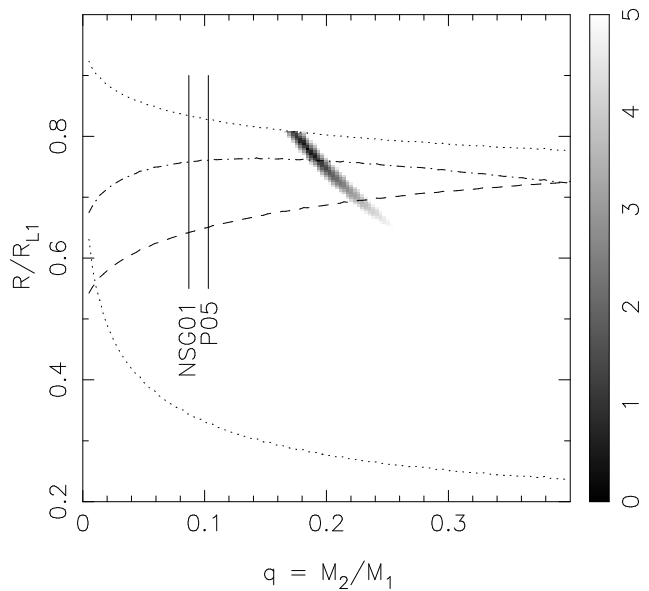}
\caption{Same as the left panel in figure \ref{massratio}, but now including the tidal truncation radius (eq.\ (\ref{truncfunc})) for an inviscid disc, indicated by the dot-dashed line.}
\label{truncation}
\end{figure}

In our derivation of the minimum mass ratio \q, we have assumed a maximum accretion disc radius equal to the minimum radius of the primary Roche lobe. It is customary to assume a slightly more stringent limit on the maximum accretion disc radius, caused by tidal truncation of the disc. See e.g.\ \citet{warnerbook} and references therein. For reference, we show in figure \ref{truncation} the often-used tidal truncation radius
\begin{equation}
\frac{R_T}{a} = \frac{0.6}{1+q}\qquad (0.03 < q < 1.0)
\label{truncfunc}
\end{equation}
for an inviscid flow, which is indeed smaller than the full minimum Roche lobe radius we assume. However, since this tidal truncation radius depends, in particular, on the unknown viscosity of the matter in the disc (and increasing with viscosity), we have not used it as a hard upper limit to the disc's radius. Interestingly, one sees that the best-fitting accretion disc radius we derive for ballistic stream velicities in the bright spot (reproduced in figure \ref{truncation}) lies very close to, but slightly above, the tidal truncation radius (\ref{truncfunc}).

The effective eccentricity value \e\ we derive for the outermost part of the disc, from the stream--disc impact spot, is smaller than typical values $e\sim0.1-0.2$ found in numerical simulations. Although the eccentricity is not one of the observables that is usually quoted in numerical studies (the resulting superhump period excess is often the only given quantity), it can be inferred that the effective eccentricity we find is significantly smaller than that found by \citet{simpsonwood} in their simulations of discs in AM CVn stars.

A possible explanation would be that tidal truncation, discussed in the previous paragraph, acts to circularise the disc at the outer edge to some extent by effectively removing particles from the most eccentric orbits there. This tidal truncation process is essentially a competition between viscosity transporting angular momentum outwards and tidal forces `dissipating' this angular momentum increasingly strongly with radius, and predictions about this tidal truncation process would thus be affected by our lack of understanding of the viscosity in accretion discs.

\subsection{Spin of the accretor and tidal synchronisation}

If the central spike originates from the surface of the accreting white dwarf, in particular if it rotates with the accretor, its width puts an interesting constraint upon the accretor's spin. The central spike is unresolved in our data, and we therefore place an upper limit of $\sim$45 km/s, equal to our formal spectral resolution, on the FWHM of the feature. For the system parameters derived for AM CVn, corotation with the orbit would imply a projected equatorial velocity of $\sim$36 km/s. The central spike we observe is thus consistent with a corotating primary, while significantly faster rotation is ruled out.

We can estimate the tidal synchronisation timescale of the accretor from this constraint. For simplicity, we assume that AM CVn has been steadily accreting matter at a rate $\dot{M_2}\sim 7\cdot 10^{-9}\,M_\odot$/yr for a long time. As long as the accretor is near corotation with the orbit, the spin-up timescale $\tau$ of the primary due to the accretion of disc matter can be written as
\begin{equation}
\tau = \frac{\omega}{\dot\omega} \approx \frac{2\pi}{P_\mathrm{orb}}k\sqrt{\frac{M_1}{G}}\frac{R_1^{3/2}}{\dot{M_2}}
\end{equation}
where $k\approx 0.18$ (see \citealt{masstransfer}) is the moment of inertia factor ($k=2/5$ for a solid sphere), and $\omega$ is the angular velocity of the accretor. Requiring that tidal synchronisation proceeds on timescales similar to or shorter than the accretion-induced spin-up of the primary, gives a maximum tidal synchronisation timescale $\tau_s \lesssim 2\cdot10^5$ yr.

A key question in binary evolution theory is whether two white dwarfs, upon Roche-lobe overflow at orbital periods of a few minutes, can manage to stabilise the accretion process and avoid a merger by feeding back angular momentum from the spun-up accretor to the orbit. This has profound implications for the number of systems that may survive the initial phase of mass transfer, when accretion proceeds at a high rate of $\dot{M_2}\sim10^{-6}M_\odot$/yr \citep{masstransfer}.

Although its absolute magnitude is highly uncertain from a theoretical point of view, tidal synchronisation is understood to scale with the orbital separation and the radius of the accretor as
\begin{equation}
\tau_s \propto \left(\frac{M_1}{M_2}\right)^2 \left(\frac{a}{R_1}\right)^6
\end{equation}
(see, e.g., \citealt{masstransfer}). This means that, for a three-minute white dwarf binary in which Roche-lobe overflow is about to commence, the tidal synchronisation timescale will be at least a factor $10^3$ shorter than for AM CVn, or $\tau_s \lesssim 200$ yr. This is an interesting result since it is exactly the regime that is needed for an appreciable fraction of binary white dwarf systems to survive the initial phase of mass transfer. See \citet{masstransfer}, in particular their figure~11.

If the central spike is thus associated with the accreting white dwarf as we think, it implies a tidal synchronisation timescale that is short enough to make the double-degenerate formation channel for AM CVn stars viable \citep{nelemans}. It would thereby also increase the viability of the direct-impact accretor scenario for the ultra-compact binaries RX J0806.3+1527 and V407 Vul \citep{directimpact}.

\section{Acknowledgments}

We thank the referee, Jan-Erik Solheim, for valuable comments and suggestions. GHAR and PJG are supported by NWO VIDI grant 639.042.201 to P.J. Groot. DS acknowledges a Smithsonian Astrophysical Observatory Clay Fellowship. GN was supported by NWO VENI grant 639.041.405 to G. Nelemans. TRM was supported by a PPARC Senior Research Fellowship. This work is based on observations made with the William Herschel Telescope operated on the island of La Palma by the Isaac Newton Group, in the Spanish Observatorio del Roque de los Muchachos of the Instituto de Astrof\'isica de Canarias. We are indebted to the astronomy staff of the Isaac Newton Group for their hospitality during our observing missions.

\end{document}